\documentclass[fleqn,11pt]{article}

\usepackage{latexsym,ifthen,epsfig,color}
\usepackage{amsmath,amssymb,amsthm}
\usepackage{float}

\newcommand{\join}{\text{\textcircled{{\footnotesize 1}}}}
\newcommand{\cojoin}{\text{\textcircled{{\footnotesize 0}}}}

\newtheorem{theorem}{Theorem}
\newtheorem{lemma}{Lemma}
\newtheorem{corollary}{Corollary}
\newtheorem{prop}{Proposition}

\newtheorem{clai}{Claim}

\begin{document}

\title{Clique Separator Decomposition of Hole- and Diamond-Free Graphs and Algorithmic Consequences}

\author{
Andreas Brandst\"adt\footnote{Fachbereich Informatik, 
Universit\"at Rostock, A.-Einstein-Str. 21, D-18051 Rostock, Germany,
{\texttt ab@informatik.uni-rostock.de}}
\and
Vassilis Giakoumakis\footnote{MIS (Mod\'elisation, Information \& Syst\`emes), Universit\'e de Picardie
Jules Verne, Amiens, France.
\texttt{e-mail: vassilis.giakoumakis@u-picardie.fr}
}
}

\maketitle

\begin{abstract}
Clique separator decomposition introduced by Tarjan and Whitesides is one of the most important graph decompositions. A graph is an {\em atom} if it has no clique separator. A {\em hole} is a chordless cycle with at least five vertices, and an {\em antihole} is the complement graph of a hole. 
A graph is {\em weakly chordal} if it is hole- and antihole-free. $K_4-e$ is also called {\em diamond}. {\em Paraglider} has five vertices four of which induce a diamond, and the fifth vertex sees exactly the two vertices of degree two in the diamond.    
In this paper we show that atoms of hole- and diamond-free graphs (of hole- and paraglider-free graphs, respectively) are either weakly chordal or of a very specific structure. Hole- and paraglider-free graphs are perfect graphs. The structure of their atoms leads to efficient algorithms for various problems.   

\medskip    

{\em Keywords}: Clique separator decomposition; hole- and diamond-free graphs; hole- and paraglider-free graphs; 
perfect graphs; efficient algorithms.
\end{abstract}

\section{Introduction, Motivation and Related Work}

A {\em clique separator} (or {\em clique cutset}) of a graph $G$ is a clique $K$ such that $G[V \setminus K]$ has more connected components than $G$. An {\em atom} is a graph without clique separator. In \cite{Tarja1985,White1984}, it is shown
that a clique separator decomposition tree of a graph can be determined in polynomial time, and in \cite{Tarja1985}, this decomposition is applied to various problems such as Minimum Fill-in, Maximum Weight Independent Set (MWIS), Maximum Weight Clique and Coloring; if the problem is solvable in polynomial time on the atoms of a hereditary graph class ${\cal C}$, it is solvable in polynomial time on class ${\cal C}$. In this paper, we are going to analyze the structure of atoms in two subclasses of hole-free graphs. 

A {\em hole} is a chordless cycle with at least five vertices, and an {\em antihole} is the complement graph of a hole. A graph is {\em hole-free} ({\em antihole-free}, respectively) if it contains no induced subgraph which is isomorphic to a hole (an antihole, respectively). 

\begin{figure}
  \begin{center}
    \epsfig{file=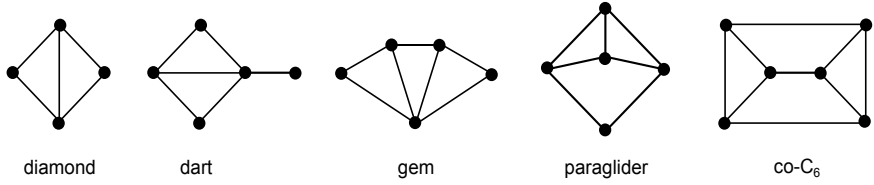}
    \caption{diamond, dart, gem, paraglider, and co-$C_6$.}
    \label{diamondetc}
  \end{center}
\end{figure}

$K_4-e$ (i.e., a clique of four vertices minus one edge) is called {\em diamond}. A {\em paraglider} has five vertices four of which induce a diamond, and the fifth vertex sees exactly the two vertices of degree two in the diamond (see Figure \ref{diamondetc}). 
Note that paraglider is the complement graph of the disjoint union $P_2 \cup P_3$ (where $P_n$ denotes a chordless path with $n$ vertices and $n-1$ edges).

\medskip
 
Cycle properties of graphs and their algorithmic aspects play a fundamental role in combinatorial optimization, discrete mathematics and computer science. Various graph classes are characterized in terms of cycle properties - among them are the classes of chordal graphs, weakly chordal graphs and perfect graphs which are of fundamental importance for algorithmic graph theory and various applications. A graph is {\em chordal} (also called {\em triangulated}) if it is hole- and $C_4$-free (where $C_4$ denotes the chordless cycle of four vertices). See e.g. \cite{BraLeSpi1999,Golum1980,McKMcM1999} for the many facets of chordal graphs. A graph is completely
decomposable by clique separator decomposition if and only if it is chordal. A graph is {\em weakly chordal} (also called {\em weakly triangulated}) if it is hole- and antihole-free. These graphs have been extensively studied in \cite{Haywa1985,HayHoaMaf1989,HaySpiSri2007,SpiSri1995}; they are perfect. In \cite{BerBorHeg2000,HaySpiSri2000}, recognition of weakly chordal graphs is solved in time ${\cal O}(m^2)$, and the MWIS problem on weakly chordal graphs is solved in time ${\cal O}(n^4)$. Chordal graphs are weakly chordal. 

\medskip

The celebrated {\em Strong Perfect Graph Theorem} ({\em SPGT}) by Chudnovsky et al. says:
\begin{theorem}[SPGT \cite{ChuRobSeyTho2006}]\label{SPGT} 
A graph is perfect if and only if it is odd-hole-free and odd-antihole-free. 
\end{theorem}

It is also well known that a graph is the line graph of a bipartite graph if and only if it is (claw,diamond,odd-hole)-free (see e.g. \cite{BraLeSpi1999}). These graphs play a fundamental role in the proof of the SPGT.   

\medskip

Since every hole $C_k$, $k \ge 7$, contains the disjoint union  of $P_2$ and $P_3$ (and the paraglider is the complement graph of $P_2 \cup P_3$), it follows that HP-free graphs are $\overline{C_k}$-free for every $k \geq 7$. Thus, by the SPGT, HP-free graphs are perfect. Our structural results for atoms of HP-free graphs, however, give a more direct way to show perfection of HP-free graphs.  

\medskip

Hole- and diamond-free graphs generalize the important class of chordal bipartite graphs (which are exactly the hole- and triangle-free graphs), and diamond-free chordal graphs are the well-known block graphs - see \cite{BraLeSpi1999} for various characterizations and the importance of chordal bipartite graphs as well as of block graphs. In \cite{BraLe2009,BraWag2010}, various characterizations of (dart,gem)-free chordal graphs are given; among others, it is shown that a graph is (dart,gem)-free chordal if and only if it results from substituting cliques into the vertices of a block graph. 

Recently there has been much work on related classes such as even-hole-free (forbidding also $C_4$) and diamond-free graphs \cite{KloMueVus2009} (see also \cite{Vusko2010}) and \cite{EscHoaSpiSri2009} dealing with the structure and recognition of $C_4$- and diamond-free graphs.  

Hole- and paraglider-free graphs obviously generalize chordal graphs. The classes of weakly chordal graphs and HP-free graphs are incomparable as the examples of paraglider (which is weakly chordal but not HP-free) and $\overline{C_6}$ (which is HP-free but not weakly chordal) show but 
HP-free graphs are closely related to weakly chordal graphs: 

\medskip

Our main result in this paper shows that atoms of hole- and paraglider-free graphs (HP-free graphs for short) are either weakly chordal or of a very simple structure close to matched co-bipartite graphs. By \cite{Tarja1985}, this has various algorithmic consequences; in section \ref{algcons}, we desribe these and others. 
 
\section{Further Basic Notions}

Let $G$ be a graph with vertex set $V(G)=V$ and edge set $E(G)=E$. Adjacency of vertices $x,y \in V$ is denoted by $xy \in E$, or $x \sim y$, or we simply say that {\em $x$ and $y$ see each other}. Nonadjacency is denoted by $xy \notin E$, or $x \not\sim y$, or {\em $x$ and $y$ miss each other}.  

The {\em open neighborhood} $N(v)$ of a vertex $v$ in $G$ is $N(v)=\{u \mid uv \in E\}$, the {\em closed neighborhood} of $v$ is $N[v]= N(v) \cup \{v\}$, and the {\em antineighborhood} $A(v)$ of $v$ is $A(v)=\{u \mid u \neq v$ and $uv \notin E\}$. 

The neighborhood $N(X)$ of a subset $X \subseteq V$ is the set of all neighbors of $x \in X$ outside $X$. For a subgraph $H$ of $G$, let $N_{H}(x)$ denote the set $N(x) \cap V(H)$ and let $N_H(X)$ denote the set $N(X) \cap V(H)$. 

For graph $G$, let $\overline {G}$ (or co-$G$) denote the complement graph of $G$, i.e., 
$\overline {G} =( V(G),\{xy \mid x \neq y $ and $x \not\sim y\})$. For $H \subseteq V$, let $G[H]$ denote the induced subgraph of $H$ in $G$. 

Let $P_{k}$ denote a chordless path with $k$ vertices $x_{1}, \ldots, x_{k}$ and edges $x_ix_{i+1}$, $1 \le i \le k-1$, and let $C_{k}$ denote a chordless cycle with the same $k$ vertices and edges $x_ix_{i+1}$, $1 \le i \le k-1$, and $x_{k}x_{1}$. 

A vertex set $U \subseteq V$ is {\em independent} if the vertices of $U$ are pairwise nonadjacent. $U$ is a {\em clique} if the vertices of $U$ are pairwise adjacent. 
Let $S_{r}$ ($K_r$, respectively) denote an independent vertex set (a clique, respectively) with $r$ vertices. 

For vertex $x$ of graph $G$ and $H \subseteq V(G)$, $x \join H$ means that $x$ is adjacent to all vertices of $H$.
In this case, we also say that $x$ is {\em total} or {\em universal} with respect to $H$. 
Correspondingly, $x \cojoin H$ means that $x$ is adjacent to no vertex of $H$. 

For $H \subseteq V(G)$ and $Q \subseteq V(G)$ with $H \cap Q = \emptyset$, $H \join Q$ means that every vertex of $H$ is adjacent to every vertex of $Q$ (we also say that $H$ and $Q$ form a {\em join})
and $H \cojoin Q$ means that no vertex of $H$ is adjacent to any vertex of $Q$ ($H$ and $Q$ form a {\em co-join}). 

Let $G$ be a graph. $G \setminus H$ or $G-H$ denotes the graph $G[V(G)-V(H)]$ induced by the set of vertices $V(G)-V(H)$. 

Let ${\cal F}$ be a set of graphs. $G$ is {\em ${\cal F}$-free} if no induced subgraph of $G$ is an element of ${\cal F}$. As already mentioned, $G$ is {\em hole-free} (is {\em antihole-free}, respectively) if no induced subgraph of $G$ is isomorphic to a hole (an antihole, respectively). 
 
A {\em co-matched bipartite graph} results from a complete bipartite graph $K_{k,k}$ by deleting a perfect matching. 
A {\em matched co-bipartite graph} is the complement of a co-matched bipartite graph, i.e., it consists of two disjoint cliques of the same size $k$, and the edges between them form a matching with $k$ edges.  

Note that $\overline{C_6}$ is a matched co-bipartite graph with six vertices. Let $A$ be a matched co-bipartite graph. Then 
$\mathop{\mathrm{left}}(A)$ denotes one of the maximal cliques of $A$ and $\mathop{\mathrm{right}}(A)$ denotes the other maximal clique of $A$. Clearly $\mathop{\mathrm{left}}(A)$ and $\mathop{\mathrm{right}}(A)$ form a bipartition of the co-matched bipartite graph $\overline{A}$ (and thus a corresponding partition of the vertex set of $A$). Subsequently, the edges between $\mathop{\mathrm{left}}(A)$ and $\mathop{\mathrm{right}}(A)$ are called {\em matching edges}. 

\section{Adjacency Properties for (Hole,Paraglider)-Free Graphs Containing $\overline{C_6}$}

In this section we describe some adjacency properties of HP-free graphs containing $\overline{C_6}$ which will be useful in the structural description of atoms of hole- and paraglider-free graphs. 

\subsection{Neighbors of $\overline{C_6}$ in HP-Free Graphs}

Throughout this section, let $G$ be an HP-free graph. As mentioned already in the introduction, the only possible antihole in an HP-free graph is $\overline{C_6}$; if $G$ is $\overline{C_6}$-free, it is weakly chordal. The following propositions are dealing with HP-free graphs containing $\overline{C_6}$. Obviously, the following holds:

\begin{prop}\label{adjcoC6} 
Pairs $x,y$ with $x \not\sim y$ in a $\overline{C_6}$ $A$ are endpoints of a $P_4$ $(x,a,b,y)$ and two $P_3$'s $(x,c,y)$, $(x,d,y)$ such that $(c,a,b,d)$ is another $P_4$ in $A$.
\end{prop}

Let $A$ be a graph isomorphic to a $\overline{C_6}$. The set of vertices outside $A$ having distance $i \ge 1$ from $A$ will be denoted by $D_i(A)$. Moreover, $D_1=D_1(A) = A_1 \cup \ldots \cup A_6$, where $A_i$, $i \in \{1,\ldots,6\}$, denotes the set of vertices outside $A$ with distance one from $A$ and having exactly $i$ neighbors in $A$ (note that $A_i$ contain only vertices which are not in $A$).

Obviously, the next property holds:

\begin{prop}\label{zadjt} 
If $x,y \in A_{1}$ with $x \sim y$, and $N_A(x)=\{t\}$, $N_A(y)=\{z\}$ with $t \neq z$ then $t \sim z$.
\end{prop}

For neighbors outside $A$ which see more than one vertex in $A$, the situation is as follows:  

\begin{prop}\label{HPfreecoC6adj}
\mbox{ }
\begin{enumerate}
\item[$(i)$] The two $A$-neighbors of any vertex in $A_2$ form an edge in $A$.
\item[$(ii)$] The three $A$-neighbors of any vertex in $A_{3}$ form a triangle in $A$.
\item[$(iii)$] $A_4=A_5=\emptyset$.
\item[$(iv)$] $A_6$ is a clique. Moreover, in a hole- and diamond-free graph, $A_6=\emptyset$. 
\item[$(v)$] If $x$ sees $A$ and $N_A(x)$ is not a clique then $x \in A_6$.   
\end{enumerate}
\end{prop}

\noindent\textbf{Proof.}
\noindent
$(i)$: If $x \in A_{2}$ sees $y$ and $z$ in $A$ with $y \not\sim z$ then by Proposition \ref{adjcoC6}, there is a $P_4$ $P$ in $A$ with endpoints $y$ and $z$. It follows that $x$ together with $P$ induce a $C_5$ in $G$, a contradiction.  

\medskip

\noindent
$(ii)$: If the neighborhood of $x \in A_{3}$ in $A$ is not a triangle then without loss of generality, $x$ sees two vertices in $\mathop{\mathrm{left}}(A)$, say $a$ and $b$, and one in $\mathop{\mathrm{right}}(A)$, say $c$. If $c$ misses $a$ and $b$ then $x,a,b,c$ together with the neighbor of $c$ in $\mathop{\mathrm{left}}(A)$ induce a paraglider, and if $c$ sees $a$ then $x,a,b,c$ together with the neighbor of $b$ in $\mathop{\mathrm{right}}(A)$ induce a paraglider - contradiction. 

\medskip

\noindent
$(iii)$: If $x \in A_4$ sees all three vertices in $\mathop{\mathrm{left}}(A)$, say $a,b,c$, and one in $\mathop{\mathrm{right}}(A)$, say $d$, then if $a$ sees $d$, $x,a,b,d$ together with the neighbor of $b$ in $\mathop{\mathrm{right}}(A)$ induce a paraglider. If $x$ sees two vertices in $\mathop{\mathrm{left}}(A)$, say $a,b$, and two vertices in $\mathop{\mathrm{right}}(A)$, say
$c,d$ then if $a$ sees $c$ and $b$ sees $d$, $x,a,d$ and the matching edge which $x$ is missing induce a $C_5$. If $a$ misses $d$ and $b$ sees $c$ then $x,a,b,c$ and the neighbor of $a$ in $\mathop{\mathrm{right}}(A)$ induce a paraglider. 

If $x \in A_5$ sees all three vertices in $\mathop{\mathrm{left}}(A)$ and two in $\mathop{\mathrm{right}}(A)$, say $d,e$, then $x,d,e$ together with the vertex $f$ which $x$ misses in $\mathop{\mathrm{right}}(A)$ and the neighbor of $f$ in $\mathop{\mathrm{left}}(A)$ induce a paraglider.     

\medskip

\noindent
$(iv)$: If there are $x,y \in A_6$ with $x \not\sim y$ then $x$ and $y$ together with any $P_1 \cup P_2$ from $A$ form a paraglider. 
Moreover, the vertices of any $P_{3}$ in $A$ together with any vertex of $A_{6}$ induce a diamond.

\medskip

\noindent
$(v)$: This property easily follows from the preceding ones.  
\qed 

\medskip

\begin{prop}\label{neighbcompar} 
Let $x \sim y$. If $x \in A_{1}$ and $y \in A_{2} \cup A_{3}$ or $x \in A_{2}$ and $y \in A_{3}$ then $N_{A}(x)$ and $N_{A}(y)$ are comparable with respect to set inclusion.
\end{prop} 

\noindent\textbf{Proof.}
As before, let $A$ be a $\overline{C_6}$, say with cliques $\mathop{\mathrm{left}}(A)=\{v_1,v_2,v_3\}$, $\mathop{\mathrm{right}}(A)=\{v_4,v_5,v_6\}$ and matching edges $v_1v_4$, 
$v_2v_5$ and $v_3v_6$. 

First let $x \in A_1$; without loss of generality, let $N_A(x)=\{v_1\}$ and assume that $y \not\sim v_1$. Recall that $y \in A_2$ or $y \in A_3$. If $\{v_2,v_3\} \subseteq N_A(y)$ then $x,y,v_1,v_2,v_3$ induce a paraglider. Thus $y$ must see at least one vertex from $\mathop{\mathrm{right}}(A)$. If $y$ sees $v_5$ then either $x,v_1,v_4,v_5,y$ or $x,v_1,v_2,v_5,y$ is a $C_5$ since by Proposition \ref{HPfreecoC6adj}, $N_A(y)$ is a clique, and similarly if $y$ sees $v_6$. Thus $y$ misses $v_5$ and $v_6$ which implies that $y$ sees $v_4$. Since by assumption, $y$ misses $v_1$, $y$ sees $v_4$ and $v_2$ or $v_3$ but this contradicts Proposition \ref{HPfreecoC6adj}.   

Now let $x \in A_2$ and $y \in A_3$; by Proposition \ref{HPfreecoC6adj}, $N_A(y)=$ $\mathop{\mathrm{left}}(A)$ or 
 $N_A(y)=$ $\mathop{\mathrm{right}}(A)$ and $N_A(x)$ is an edge in $A$. If $N_A(x)=\{v_1,v_2\}$ and $N_A(x)$ and $N_A(y)$ are not comparable then $N_A(y)=$ $\mathop{\mathrm{right}}(A)$ but now $x,v_2,v_3,v_6,y$ is a $C_5$ - contradiction. If however $N_A(x)=\{v_1,v_4\}$ and without loss of generality, $N_A(y)=$ $\mathop{\mathrm{left}}(A)$ then $x,y,v_2,v_5,v_4$ is a $C_5$ which shows Proposition \ref{neighbcompar}.
\qed

\begin{prop}\label{Nxyclique} 
For all $x,y \in A_{2}$ with $x \sim y$, $N_{A}(x)\cup N_{A}(y)$ is a clique.
\end{prop}

\noindent\textbf{Proof.}
By Proposition \ref{HPfreecoC6adj}, $N_{A}(x)$ and $N_{A}(y)$ are edges. Assume to the contrary that there are $z \in N_A(x)$ and $t \in N_A(y)$ with $z \not\sim t$. Thus $z \notin N_A(y)$ and $t \notin N_A(x)$. By Proposition \ref{adjcoC6}, there is a $P_4$ $(z,u,v,t)$ in $A$. Since $N_A(x)$ is an edge, $x$ misses $v$, and likewise $y$ misses $u$. To avoid a hole in the subgraph induced by $\{x,z,u,v,t,y\}$, we obtain
$x \sim u$ and $y \sim v$ which implies that $N_{A}(x) \cup N_A(y) = \{z,u,v,t\}$. Then by Proposition \ref{adjcoC6} there is a $P_{3}$ $(z,w,t)$ 
in $A$ such that $x$ and $y$ miss $w$ and consequently $x,z,w,t,y$ induce a $C_{5}$ in $G$, a contradiction. 
\qed

\medskip

Now it is easy to see that by Propositions \ref{zadjt}, \ref{HPfreecoC6adj}, \ref{neighbcompar}, and \ref{Nxyclique}, we obtain: 

\begin{corollary}\label{coroNxyclique}
For all $x,y \in D_1$ with $x \sim y$ and at least one of $x,y$ does not belong to $A_{3}$,  $N_{A}(x) \cup N_{A}(y)$ is a clique. 
\end{corollary}
 
\begin{prop}\label{chainNxycompar} 
Let $x,y \in D_1$ with $x \not\sim y$ be the endpoints of a chordless path $P$ whose internal vertices do not belong to $D_1 \cup A$. Then 
\begin{itemize}
\item[$(i)$] $P$ contains exactly three vertices $x,w,y$ and 
\item[$(ii)$] $N_{A}(x)$ and $N_{A}(y)$ are comparable. 
\end{itemize}
\end{prop}

\noindent\textbf{Proof.}
$(i)$: Assume to the contrary that $P$ contains at least four vertices. Let $u$ and $v$ be two vertices of $A$ such that $u\in N_{A}(x)$ and $v\in N_{A}(y)$ and let $Q$ be a chordless path in $A$ joining $u$ and $v$ (possibly length$(Q)=0$, i.e.,
$u = v$). Now it is easy to verify that the graph induced by the vertices of $P \cup Q$ contains a hole, a contradiction. 

\medskip

\noindent
$(ii)$: Assume to the contrary that $N_{A}(x)$ and $N_{A}(y)$ are not comparable. Let $z$ and $t$ be two vertices of $A$ such that $z \in N_{A}(x)-N_{A}(y)$ and $t\in N_{A}(y)-N_{A}(x)$. If $z$ is adjacent to $t$ then $x,z,t,y,w$ (where $w$ is the vertex from condition $(i)$) induce a $C_{5}$.
Hence $z \not\sim t$, and by Proposition \ref{adjcoC6}, there is a $P_{4}$ $(z,a,b,t)$ in $A$. Since by Proposition \ref{HPfreecoC6adj}, $N_{A}(x)$ and $N_{A}(y)$ are cliques, neither $x$ nor $y$ can be adjacent to both vertices $a$ and $b$. It follows that the subgraph induced by $x,z,a,b,t,y,w$
 contains a hole, a contradiction. 
\qed 

\begin{prop}\label{A6total} 
Let $A^{*}$ be a maximal matched co-bipartite subgraph of $G$ containing $A$. Then the following hold:
\begin{itemize}
\item[$(i)$] Every vertex of $A_6$ is total with respect to $V(A^{*})$.
\item[$(ii)$] If $x$ and $y$ are vertices of $G \setminus A^{*}$ with $x,y \in A_{3}$, $N_{A}(x)=\mathop{\mathrm{left}}(A)$ and $N_{A}(y)=\mathop{\mathrm{right}}(A)$ then $x \not\sim y$. 
\end{itemize}
\end{prop}

\noindent\textbf{Proof.} 
$(i)$: Assume to the contrary that for some $x \in A_{6}$ and $y \in V(A^{*})-V(A)$, $x \not\sim y$ holds. Assume without loss of generality that $y \in \mathop{\mathrm{left}}(A^{*})$ and let $z$ be the neighbor of $y$ in $\mathop{\mathrm{right}}(A^{*})$. Consider the subgraph $H$ of $G$ induced by $a,b,c,d,y,z$ where $a,b,c,d$ are four vertices of $A$ forming a $C_{4}$. Clearly, $H$ is isomorphic to a $\overline{C_6}$. Since $x$ is total with respect to $\{a,b,c,d\}$, $x$ will be adjacent to four or five vertices of $H$ and we obtain a contradiction to Proposition \ref{HPfreecoC6adj}. 

\medskip

\noindent
$(ii)$: First observe that if $A^{*}=A$ then $x\not\sim y$ 
for otherwise the graph induced by $V(A)\cup\{x,y\}$ is a matched
co-bipartite graph and this contradicts the maximality of $A^{*}$.
Thus, we can suppose that $V(A^{*})-V(A)\neq\emptyset$. 

Assume to the contrary that $x \sim y$ and consider any edge $zt$ of $A^{*}-A$ such that $z \in \mathop{\mathrm{left}}(A^*)$ and $t \in \mathop{\mathrm{right}}(A^*)$. Let $Q$ be the graph induced by $z,t$ and four vertices $a,b,c,d$ forming a $C_{4}$ in $A$ such that $\{a,b\} \subset \mathop{\mathrm{left}}(A)$ and $\{c,d\}\subset \mathop{\mathrm{right}}(A)$.
Clearly $Q$ is isomorphic to a $\overline{C_6}$. 

We shall prove that $x\sim z$, $y\sim t$, $x\not\sim t$ and $y\not\sim z$. 
Observe first that since $x$ misses $c,d$ and $y$ misses $a,b$, we must have that $x\not\sim t$ and $y\not\sim z$ for otherwise $N_{Q}(x)$ or $N_{Q}(y)$ would not be a clique which contradicts Proposition \ref{HPfreecoC6adj}. 

Let $Q_2$ ($Q_3$, respectively) denote the vertices outside $Q$ having exactly two neighbors (three neighbors, respectively) in $Q$. 
Now $x\sim z$ and $y\sim t$ for otherwise since $x$ sees $a$ and $b$, and $y$ sees $c$ and $d$, we would have $x\in Q_{2}$ and $y\in Q_{2}\cup Q_{3}$ or $x\in Q_{2}\cup Q_{3}$ and $y\in Q_{2}$, and we obtain a contradiction to Proposition \ref{neighbcompar} or Proposition \ref{Nxyclique}. Hence $x \join \mathop{\mathrm{left}}(A^{*})$, 
$x \cojoin \mathop{\mathrm{right}}(A^{*})$, $y \join \mathop{\mathrm{right}}(A^{*})$ and 
$y \cojoin \mathop{\mathrm{left}}(A^*)$ and consequently $V(A^{*}) \cup \{x,y\}$ induces a graph isomorphic to a matched co-bipartite graph which contradicts to the assumed maximality of $A^{*}$. 
\qed  

\subsection{A Lemma for Atoms of HP-Free Graphs}

The subsequent Lemma \ref{A12xyempty} describes an essential property of HP-free atoms which will lead to a structural description of HP-free graphs.  

\medskip

Let $G$ be an HP-free graph, let $A$ be an induced $\overline{C_6}$ in $G$ and let $xy$ be a matching edge of $A$ with $x \in \mathop{\mathrm{left}}(A)$ and $y \in \mathop{\mathrm{right}}(A)$. We use the following notation: 

\begin{itemize}
\item $A_{2}[xy]:=\{u \mid u \in A_{2}, N_{A}(u)=\{x,y\}\}$
\item $A_{1}[xy]:=\{uv \in E \mid u,v \in A_{1}, N_{A}(u)=\{x\}, N_{A}(v)=\{y\}\}$. 
\end{itemize}

By $V(A_{1}[xy])$, we denote the set of vertices in $A_{1}[xy]$.  

\begin{lemma}\label{A12xyempty} 
In an HP-free atom, $A_{1}[xy]=A_{2}[xy]=\emptyset$. 
\end{lemma}

\noindent\textbf{Proof.}
Assume to the contrary that at least one of the two sets is nonempty.
Recall that by Proposition \ref{HPfreecoC6adj} $(iv)$, $A_{6}$ is a clique which implies that $\{x,y\}\cup A_{6}$ is a clique.
Let $G':=G \setminus (\{x,y\} \cup A_{6})$ and $A':=A \setminus \{x,y\}$. Clearly the vertices of $A'$ form a $C_{4}$, say $C=(a,b,c,d)$ with $\mathop{\mathrm{left}}(A)=\{x,a,d\}$ and $\mathop{\mathrm{right}}(A)=\{y,b,c\}$.  Since $G$ is an atom, $\{x,y\}\cup A_{6}$ can not be a clique cutset and consequently, $G'$ contains a path between some vertex $x_0 \in A_{2}[xy] \cup V(A_{1}[xy])$ and $x_k \in A'$. 
Let $L=(x_{0},x_{1},\ldots,x_{k})$ be such a path of minimum length in $G'$.
If $x_{0}y_0 \in A_{1}[xy]$ then we assume without loss of generality that $x_{0} \sim x$ and $y_{0} \sim y$. 

\begin{clai}\label{length(l)>2}  
length$(L)>2$.
\end{clai}

\noindent{\em Proof of Claim $\ref{length(l)>2}$.} Assume not - then $L=(x_0,x_1,x_2)$ with $x_2 \in A'$. 

\medskip

\noindent 
Assume first that $x_{0} \in A_{2}[xy]$. Since by Proposition \ref{HPfreecoC6adj}, $N_{A}(x_{1})$ is a clique (recall that $x_1 \notin A_6$) and $N_{A}(x_{1}) \cap \{a,b,c,d\} \neq\emptyset$, if $x_{1}\in A_{1}\cup A_{3}$
then $N_{A}(x_{0})$ is not comparable with $N_{A}(x_{1})$ which
contradicts Proposition \ref{neighbcompar} and if $x_{1}\in A_{2}$, $N_{A}(x_{0})\cup N_{A}(x_{1})$
is not a clique which contradicts Proposition \ref{Nxyclique}. 

\medskip

\noindent 
Assume now that $x_{0}\in V(A_{1}[xy])$ (recall that we assumed $x_0 \sim x$). By Proposition \ref{zadjt} and Proposition \ref{neighbcompar} we deduce that $N_{A}(x_{1}) \subseteq \{x,a,d\}$ and that $y_{0} \not\sim x_{1}$. 
Let $u$ be a neighbor of $x_{1}$ in $\{a,d\}$ and $v$ the vertex of $\{b,c\}$
adjacent to $u$. Then $x_{0},x_{1},u,v,y,y_{0}$ induce a $C_6$, a contradiction which shows Claim \ref{length(l)>2}. 
$\Diamond$ 

\medskip

Since length$(L)$ is assumed to be minimum, none of $x_{1},\ldots,x_{k-2}$ can be in $A_{2} \cup A_{3} \cup V(A_{1}[xy]) \cup A_{2}[xy]$. It follows that if a vertex $x_{i} \in \{x_{1},\ldots,x_{k-2}\}$ belongs to $D_{1}$ then $x_{i}\in A_1 - V(A_1[xy])$. Let 
$$Q:= \{x_{1},\ldots,x_{k-2}\} \cap (A_1 - V(A_1[xy])).$$ 
 
\begin{clai}\label{Qneqempty} 
If $x_0 \in A_{2}[xy]$ then $Q \neq \emptyset$.
\end{clai}

\noindent{\em Proof of Claim $\ref{Qneqempty}$.} Assume $Q = \emptyset$; then none of $x_{1},\ldots,x_{k-2}$
belongs to $D_{1}$ and consequently by Proposition \ref{chainNxycompar},
$N_{A}(x_{k-1})$ and $N_{A}(x_{0})=\{x,y\}$ must be comparable. 
By Proposition \ref{HPfreecoC6adj}, $N_{A}(x_{k-1})$ must be a clique (recall that $x_k \in \{a,b,c,d\}$, and since the path in $G'$ contains no vertex from $A_6$, we have $x_{k-1} \notin A_{6}$). Thus we obtain a contradiction which shows Claim \ref{Qneqempty}. 
$\Diamond$

\begin{clai}\label{NAQ} 
If $Q \neq\emptyset$ then either $N_A(Q)=\{x\}$ or $N_A(Q)=\{y\}$.
\end{clai}

\noindent{\em Proof of Claim $\ref{NAQ}$.} Assume not; then there are two vertices $x_{i}$
and $x_{j}$ in $Q$, $1\leq i<j\leq k-2$, such that $N_{A}(x_{i})\neq N_{A}(x_{j})$
and for all $k$, $i<k<j$, $x_{k}\notin D_{1}$. Observe that $j>i+1$ for
otherwise $x_i$ would be adjacent to $x_j$ and consequently
$x_i$ and $x_j$ would belong to $V(A_1[x,y])$, a contradiction.
Now $N_{A}(x_{i})$ and $N_{A}(x_{j})$ are not comparable - a contradiction to Proposition \ref{chainNxycompar} which shows Claim \ref{NAQ}. 
$\Diamond$

\begin{clai}\label{Nxk-1sub} 
If $Q \neq\emptyset$ then $N_{A}(Q)=\{x\}$ implies that $N(x_{k-1})\subseteq \mathop{\mathrm{left}}(A)$ and $N_{A}(Q)=\{y\}$ implies that $N(x_{k-1})\subseteq \mathop{\mathrm{right}}(A)$.
\end{clai}

\noindent{\em Proof of Claim $\ref{Nxk-1sub}$.} Let $x_{s}$, $1\leq s\leq k-2$, be a vertex of path $L$ with $x_s \in Q$ such that $s$ is maximum with respect to these properties.

Assume first that $x_{k-1}\in A_{1}$. Then $x_{s} \sim x_{k-1}$ for otherwise, by Proposition \ref{chainNxycompar}, $N_{A}(x_{k-1})$ must be comparable with $N_{A}(x_{s})$ and we obtain a contradiction to the fact that $x_{k-1}$ has a neighbor in $\{a,b,c,d\}$. Proposition \ref{zadjt} implies that $N_{A}(x_{k-1})\sim N_{A}(x_{s})$ and consequently $N_{A}(x_{k-1})$ is contained either in $\{a,d\}\subset \mathop{\mathrm{left}}(A)$ if $N_{A}(x_{s})=\{x\}$ or in $\{b,c\}\subset \mathop{\mathrm{right}}(A)$ if $N_{A}(x_{s})=\{y\}$. 

Now assume that $x_{k-1} \in A_{2}\cup A_{3}$. Then Proposition \ref{neighbcompar} and Proposition \ref{chainNxycompar} imply that $N_{A}(x_{k-1})$ and $N_{A}(x_{s})$ must be comparable. Claim \ref{Nxk-1sub} follows from the fact that $N_{A}(x_{k-1})$ is a clique and at least one of the vertices of $\{a,b,c,d\}$ belongs to $N_{A}(x_{k-1})$. 
$\Diamond$

\begin{clai}\label{NAQxk-1} 
For $x_{0}\in V(A_1[xy])$, the following hold:
\begin{enumerate}
\item[$(i)$] If $Q \neq\emptyset$ then $N_{A}(Q)=\{x\}$.
\item[$(ii)$] $N_{A}(x_{k-1}) \subseteq \mathop{\mathrm{left}}(A)$.
\end{enumerate}
\end{clai}

\noindent{\em Proof of Claim $\ref{NAQxk-1}$.} 
$(i)$: Recall that for $x_0 \in V(A_1[xy])$, we assumed that $N_A(x_0)=\{x\}$. 
Let $x_{i}$ be a vertex such that $x_{i} \in Q$ and $i$ is as small as possible. Recall that by Claim \ref{NAQ}, either $N_{A}(Q)=\{x\}$ or $N_{A}(Q)=\{y\}$ holds.

If $i=1$ and $N_{A}(Q)=\{y\}$ then $x_1 \in V(A_{1}[xy])$ since $x_1 \sim x_0$ - a contradiction to the fact that every vertex of $Q$ belongs to $A_{1}-V(A_{1}[xy])$. Thus, $N_{A}(x_{1})=\{x\}$ and also $N_{A}(Q)=\{x\}$. 

If $i>1$ then $x_{1}\in D_{2}$ and by Proposition \ref{chainNxycompar} we obtain that $i=2$ and $N_{A}(x_{2})=\{x\}$. Then by Claim \ref{Qneqempty} we obtain that $N_{A}(Q)=\{x\}$ as claimed.

\medskip

\noindent
$(ii)$: If $Q \neq\emptyset$ then $N_{A}(x_{k-1}) \subseteq \mathop{\mathrm{left}}(A)$ follows by the fact that $N_{A}(Q)=\{x\}$ and Claim \ref{Nxk-1sub}. In the other case, if $Q=\emptyset$ then no vertex of $\{x_{1},\ldots,x_{k-2}\}$ is in $D_{1}$. Proposition \ref{chainNxycompar} implies that $N_{A}(x_{k-1})$ and $N_{A}(x_{0})$ must be comparable, and since by assumption $N_{A}(x_{0})=\{x\}$ and $N_{A}(x_{k-1})$ is a clique, we obtain Claim \ref{NAQxk-1}.
$\Diamond$

\medskip

Let $u \in \{a,d\}$ be a neighbor of $x_{k-1}$ and let $v$ be the neighbor of $u$ in $\mathop{\mathrm{right}}(A)$ which clearly is different from the vertex $y$. If $x_{0}\in A_{2}[xy]$ then by Claim \ref{Qneqempty}, $Q \neq\emptyset$ and by Claim \ref{NAQ}, $N_{A}(Q)=\{x\}$ or $N_{A}(Q)=\{y\}$. Assume without loss of generality that $N_{A}(Q)=\{x\}$; then by Claim \ref{Nxk-1sub}, we have $N(x_{k-1})\subseteq \mathop{\mathrm{left}}(A)$. Then the subgraph induced by $x_{0},\ldots,x_{k-1},u,v,y$ is a hole, a contradiction. Hence $x_{0} \in V(A_{1}[xy])$.
By Claim \ref{NAQxk-1}, if $Q \neq\emptyset$ then $N_{A}(Q)=\{x\}$. It follows that the subgraph induced by $x_{0},\ldots,x_{k-1},u,v,y,y_{0}$ is a hole, a contradiction which shows Lemma \ref{A12xyempty}. 
\qed 

\section{Structure of (Hole,Paraglider)-Free and (Hole,Diamond)-Free Atoms}

Recall that HP-free (HD-free, respectively) denotes hole- and paraglider-free (hole- and diamond-free, respectively).   

\begin{theorem}\label{HPatomcobip}
If $G$ is an HP-free atom containing an induced $\overline{C_6}$ $A$, and $A_6$ denotes the set of vertices which are universal for $A$ then $G \setminus A_{6}$ is a matched co-bipartite graph. 
\end{theorem}

\noindent\textbf{Proof.} 
Assume the contrary; let $G':=G\setminus A_{6}$ and let $A^{*}$ be a maximal matched co-bipartite subgraph in $G'$ containing $A$. Let $W:=V(G')-V(A^{*})$; by assumption, $W \neq \emptyset$.
We define a partition $\pi(W)$ of the vertices of $W$ according to their distance from $A^{*}$: $W=W_{1} \cup \ldots \cup W_{k}$ where $W_{i}:=\{x\in W \mid d(x,A^{*})=i\}$, $i=1,\ldots,k$. Thus, $W_1 = (W \cap (A_{1} \cup A_{2} \cup A_{3})) \cup (W \cap D^*_2)$ where $D^*_2$ denotes the set of vertices which are in distance two from $A$ and which see a vertex in $A^*$.   
The vertices in $W_{2}$ have distance at least two from $A$. 

\begin{clai}\label{noninboth} 
No vertex in $W_{1}$ has neighbors in both $\mathop{\mathrm{left}}(A^{*})$ and $\mathop{\mathrm{right}}(A^{*})$.
\end{clai}

\noindent{\em Proof of Claim $\ref{noninboth}$.} 
Assume to the contrary that for some $x \in W_{1}$, there are $y$ and $z$ with $y \in \mathop{\mathrm{left}}(A^{*})$ and $z \in \mathop{\mathrm{right}}(A^{*})$ such that $x \sim y$ and $x \sim z$. 

Suppose first that $y \sim z$. Consider the graph $Q$ induced by $y,z$ and four vertices $a,b,c,d$ of $A$ forming a $C_{4}$ such that $\{y,z\}\cap\{a,b,c,d\}=\emptyset$. Clearly $Q$ is isomorphic to a $\overline{C_6}$. Then since by Lemma \ref{A12xyempty}, $Q_2[yz]=\emptyset$ (where as before, $Q_2[yz]$ denotes the vertices outside $Q$ seeing exactly $y$ and $z$ in $Q$), $x$ can not belong to $D_{2}(A)$ and consequently $N(x)\cap\{a,b,c,d\}\neq\emptyset$, that is, $x \in A_{1}\cup A_{2}\cup A_{3}$. Since by Proposition \ref{HPfreecoC6adj}, $N_{Q}(x)$ is a clique and by assumption $x$ sees both $y$ and $z$, we obtain a contradiction. 

Now suppose that $y\not\sim z$ and consider the graph $H$ induced by $y,z,y_{1},z_{1},a,b$ where $y_{1}$ is the neighbor of $y$ in $\mathop{\mathrm{right}}(A^{*})$, $z_{1}$ is the neighbor of $z$ in $\mathop{\mathrm{left}}(A^*)$, $ab$ is any edge of $A$ such that $a\in \mathop{\mathrm{left}}(A)$, $b \in \mathop{\mathrm{right}}(A)$ and $\{a,b\}\cap\{y,y_{1},z,z_{1}\}=\emptyset$. Clearly $H$ is isomorphic to a $\overline{C_6}$. Since by assumption $x$ sees both $y$ and $z$, $N_{H}(x)$ is not a clique which by Proposition \ref{HPfreecoC6adj} $(v)$ implies that $x$ sees all vertices of $H$ and thus also $x \sim a$ and $x \sim b$ with $a \in \mathop{\mathrm{left}}(A)$ and $b \in \mathop{\mathrm{right}}(A)$. Since by Proposition \ref{HPfreecoC6adj}, $x \not\in A_3$, by Lemma \ref{A12xyempty}, $x \not\in A_2[a,b]$ and by assumption, $x \not\in A_{6}$, we obtain a contradiction.
$\Diamond$

\medskip

\noindent
We define now the following sets: 
\begin{itemize}
\item[ ] $\mathop{\mathrm{left}}(W_{1}):=\{x\in W_{1}\mid N_{A^{*}}(x)\subseteq \mathop{\mathrm{left}}(A^{*})\}$ and 
\item[ ] $\mathop{\mathrm{right}}(W_{1}):=\{x\in W_{1}\mid N_{A^{*}}(x) \subseteq \mathop{\mathrm{right}}(A^{*})\}.$ 
\end{itemize}

\noindent
By Claim \ref{noninboth}, $\mathop{\mathrm{left}}(W_1) \cap \mathop{\mathrm{right}}(W_1) = \emptyset$. Thus $W_{1} = $ $\mathop{\mathrm{left}}(W_{1}) \cup \mathop{\mathrm{right}}(W_{1})$ is a partition of $W_1$. 

\begin{clai}\label{nobetwlar} 
There is no edge between $\mathop{\mathrm{left}}(W_{1})$ and $\mathop{\mathrm{right}}(W_{1})$.
\end{clai}

\noindent{\em Proof of Claim $\ref{nobetwlar}$.} 

Assume to the contrary that $x \sim y$ for some $x \in \mathop{\mathrm{left}}(W_{1})$ and $y \in \mathop{\mathrm{right}}(W_{1})$. Recall that $D_1$ denotes the vertices in distance one to $A$. 
We first show: 
\begin{equation}\label{AAA}
x \mbox{ and } y \mbox{ can not be both in } D_1. 
\end{equation}

Assume to the contrary that $x,y \in D_1$. Then by Proposition \ref{A6total} $(ii)$, $x,y \in A_{3}$
is impossible. Suppose without loss of generality that $x \notin A_{3}$, i.e., $x \in A_{1} \cup A_{2}$ and $y \in A_{1} \cup A_{2} \cup A_{3}$. If $x \in A_{1}$ and $y \in A_{2} \cup A_{3}$ or $x \in A_{2}$ and
$y \in A_{1} \cup A_{3}$, Proposition \ref{neighbcompar} implies that $N_{A}(x)$
and $N_{A}(y)$ are comparable, and if $x,y \in A_{2}$, Proposition \ref{Nxyclique} implies that $N_{A}(x)\cup N_{A}(y)$ is a clique. But since $x \in \mathop{\mathrm{left}}(W_{1})$ and $y \in \mathop{\mathrm{right}}(W_{1})$,
none of these cases can occur. It follows that $x,y \in A_{1}$. However, by Lemma \ref{A12xyempty},
such a pair of adjacent vertices can not exist, a contradiction.
$\diamond$ 

\medskip

It follows that at least one of $x$ or $y$ is in $D_{2}$. Assume that $x \in D_{2}$ and let $u$ be a neighbor
of $x$ in $D_{1}$. Suppose first that also $y \in D_{2}$ and let $v$ be a neighbor of $y$ in $D_{1}$. Obviously  
$u \in \mathop{\mathrm{left}}(W_{1})$ and $v \in \mathop{\mathrm{right}}(W_{1})$.
Since by assumption $x,y \in D_{2}$, Proposition \ref{chainNxycompar} $(i)$ implies that $u \sim v$ and we obtain a contradiction with (\ref{AAA}). Consequently, $y \in D_{1}$. Since $N_{A}(u)$ and $N_{A}(y)$ are not comparable, Proposition \ref{chainNxycompar} $(ii)$ implies that $u \sim y$ and again we obtain a contradiction with (\ref{AAA}). This shows Claim $\ref{nobetwlar}$. $\Diamond$ 

\medskip

For the partition $\pi(W)=\{W_{1},\ldots,W_{k}\}$, $k \geq 1$, define the following sets for every $i \in \{2,\ldots,k\}$: 

\begin{itemize}
\item[ ] $\mathop{\mathrm{left}}(W_{i}):=\{x\in W_{i} \mid \exists y\in \mathop{\mathrm{left}}(W_{i-1}$) such that $x\sim y\}$ and 
\item[ ] $\mathop{\mathrm{right}}(W_{i}):=\{x\in W_{i} \mid \exists y\in \mathop{\mathrm{right}}(W_{i-1})$ such that $x\sim y\}$. 
\end{itemize}

\begin{clai}\label{leftright} 
$($$\mathop{\mathrm{left}}$$(W_1) \cup \ldots \cup \mathop{\mathrm{left}}(W_k)) \cap (\mathop{\mathrm{right}}(W_1) \cup \ldots \cup \mathop{\mathrm{right}}(W_k)) = \emptyset$
and $(\mathop{\mathrm{left}}(W_1)\cup \ldots \cup \mathop{\mathrm{left}}(W_{k})) \cojoin (\mathop{\mathrm{right}}(W_1)\cup \ldots \cup \mathop{\mathrm{right}}(W_{k}))$.
\end{clai}

\noindent{\em Proof of Claim $\ref{leftright}$.} We shall prove the claim by induction on $k$. By
Claims \ref{noninboth} and \ref{nobetwlar}, the result is true for $k=1$. By the induction
hypothesis the result is true for $k<s$, $s>1$. Assume to the contrary that the result is false  
for $W_{s}\in\pi(W)$. Then there must be a chordless path $L_{1}=(x_{1}, \ldots, x_{s-1},x, y_{s-1}, \ldots ,y_{1})$ or a chordless path $L_{2}=(x_{1},\ldots, x_{s-1},x, y, y_{s-1},\ldots, y_{1})$ such that $x_{i}\in \mathop{\mathrm{left}}(W_{i})$, $y_{i}\in \mathop{\mathrm{right}}(W_{i})$, $i \in \{1,\ldots,s-1\}$ and $x,y\in W_{s}$. By the induction hypothesis there is no edge between $\{x_{1},\ldots,x_{s-1}\}$ and $\{y_{1},\ldots,y_{s-1}\}$. Let
$L=(x_{1},z_{1},\ldots,z_{r},y_1)$, $r\geq 2$, be a chordless path joining $x_1$ and $y_1$ such that $z_{i}\in A^{*}$, $i \in \{1,\ldots,r\}$, which clearly exists. It is easy to see that the graph induced by the vertices of $L_{1}$ and $L$ or by the vertices of $L_{2}$ and $L$ is isomorphic to a hole - a contradiction. This shows Claim \ref{leftright}.
$\Diamond$

\medskip

\noindent
Let 
\begin{itemize}
\item[ ] $\mathop{\mathrm{left}}(W):=(\mathop{\mathrm{left}}(W_{1})\cup \ldots \cup \mathop{\mathrm{left}}(W_{k}))$ and 
\item[ ] $\mathop{\mathrm{right}}(W):=(\mathop{\mathrm{right}}(W_{1})\cup \ldots \cup \mathop{\mathrm{right}}(W_{k})$.
\end{itemize} 
 
\noindent 
By Claim \ref{leftright}, $\mathop{\mathrm{left}}(W)$ and $\mathop{\mathrm{right}}(W)$ form a partition of $W$.

\begin{clai}\label{leftrightcojoin} 
$\mathop{\mathrm{left}}(W) \cojoin \mathop{\mathrm{right}}(W) \cup \mathop{\mathrm{right}}(A^{*})$ and 
$\mathop{\mathrm{right}}(W) \cojoin \mathop{\mathrm{left}}(W) \cup \mathop{\mathrm{left}}(A^{*})$. 
\end{clai}

\noindent{\em Proof of Claim $\ref{leftrightcojoin}$.} Indeed, by Claim \ref{leftright}, we have that $\mathop{\mathrm{left}}(W) \cojoin \mathop{\mathrm{right}}(W)$. By Claim \ref{noninboth}, we have that $\mathop{\mathrm{left}}(W_{1}) \cojoin \mathop{\mathrm{right}}(A^{*})$ and $\mathop{\mathrm{right}}(W_{1}) \cojoin \mathop{\mathrm{left}}(A^{*})$, and by the construction of $W_{2},\ldots,W_{k}$ we have that $(W_{2} \cup \ldots \cup W_{k})\cojoin V(A^{*})$. This shows Claim \ref{leftrightcojoin}.
$\Diamond$

\medskip
 
Since by assumption $G'=G \setminus A_{6}$ is not isomorphic to a matched co-bipartite graph, we must have that $W\neq\emptyset$. Assume without loss of generality that $\mathop{\mathrm{left}}(W) \neq\emptyset$. 
Then since by Proposition \ref{A6total} $(i)$, $A_{6}\cup$ $\mathop{\mathrm{left}}(A^{*})$ is a clique and since by Claim \ref{leftrightcojoin}, there is no edge between $\mathop{\mathrm{left}}(W)$ and $\mathop{\mathrm{right}}(W) \cup \mathop{\mathrm{right}}(A^{*})$, $A_{6} \cup \mathop{\mathrm{left}}(A^{*})$ would be a clique cutset in $G$ which contradicts our assumption that $G$ is an atom. This finishes the proof of Theorem \ref{HPatomcobip}. 
\qed

\begin{corollary}\label{HPstructure}
Let $G$ be a $($hole,paraglider$)$-free graph. 
\begin{itemize}
\item[$(i)$] If $G$ is $\overline{C_6}$-free then $G$ is weakly chordal.  
\item[$(ii)$] If $G$ is an atom containing an induced $\overline{C_6}$ then $G$ is the join of a matched co-bipartite graph and a clique.
\end{itemize} 
\end{corollary}

\noindent\textbf{Proof.} 
$(i)$: Recall that $HP$-free graphs are $\overline{C_k}$-free for $k \ge 7$.

\medskip

\noindent 
$(ii)$: Indeed by Theorem \ref{HPatomcobip}, for a $\overline{C_6}$ $A$ in $G$, $G'=G\setminus A_{6}$ is a matched co-bipartite graph. By Proposition \ref{A6total}, $A_{6} \join V(G')$, and by Proposition \ref{HPfreecoC6adj}, $A_6$ is a clique. 
\qed 

\medskip

Since by Proposition \ref{HPfreecoC6adj} $(iv)$, in (hole,diamond)-free graphs $A_6=\emptyset$, we have:

\begin{corollary}\label{HDstructure}
Let $G$ be a $($hole,diamond$)$-free graph. 
\begin{itemize}
\item[$(i)$] If $G$ is $\overline{C_6}$-free then $G$ is weakly chordal.  
\item[$(ii)$] If $G$ is an atom containing an induced $\overline{C_6}$ then $G$ is a matched co-bipartite graph.
\end{itemize} 
\end{corollary}

\section{Algorithmic Consequences}\label{algcons}

In \cite{Tarja1985}, for various problems such as Minimum Fill-in, Maximum Independent Set, Maximum Clique and Coloring, it is shown that whenever these problems are efficiently solvable on the atoms of a graph class, they are efficiently solvable on the graphs of the class.   
For perfect graphs, Maximum Independent Set, Maximum Clique and Coloring are known to be solvable in polynomial time 
\cite{GroLovSch1981,GroLovSch1984} using the ellipsoid method (but from a practical point of view, this is not an efficient solution of the problems). 

(Hole,paraglider)-free graphs are perfect as the Strong Perfect Graph Theorem implies (a more direct way
can use Theorem \ref{HPatomcobip} and Corollary \ref{HPstructure} and the fact that a graph is perfect if its atoms are perfect). 

The clique separator approach gives direct combinatorial algorithms for the problems mentioned above: 

Recognition of weakly chordal graphs can be done in ${\cal O}(m^2)$ \cite{BerBorHeg2000,HaySpiSri2000}, and recognition of matched co-bipartite graphs can be easily done in linear time. Thus, given an input graph, determine its atoms and check whether they are either weakly chordal or are the join of a clique and a matched co-bipartite graph. If not then the input graph is not (hole,paraglider)-free. Otherwise solve the problems on the atoms and finally combine the solutions as described in \cite{Tarja1985}.

For matched co-bipartite graphs, MWIS is trivial. A first polynomial time algorithm for weakly chordal graphs is given in \cite{HayHoaMaf1989}, and in \cite{SpiSri1995}, MWIS is solved in time ${\cal O}(n^4)$ for weakly chordal graphs. Thus, the time bound for MWIS on HP-free graphs is roughly ${\cal O}(n^6)$: Determine whether the input graph is weakly chordal. If yes, use the algorithm for weakly chordal graphs. If not, check whether all prime atoms are matched co-bipartite, and if yes, then use the trivial algorithm for these graphs. If not, the input graph is not HP-free.  

For Maximum Clique and Coloring one can proceed in a similar way. For Maximum Clique on diamond-free graphs, however, there is a more direct way to solve the problem efficiently by switching to the complement graph and the complement problem MWIS: If $G$ is gem-free (see Figure \ref{diamondetc} for gem) then $\overline{G}$ has the property that for every vertex, its antineighborhood is $P_4$-free, i.e., a cograph. This means that one can solve the MWIS problem for such graphs in time ${\cal O}(n m)$ in the obvious way.     

In \cite{BouTod2001}, a ${\cal O}(n^6)$ algorithm is given for Minimum Fill-In on weakly chordal graphs. Minimum Fill-In on matched co-bipartite graphs is efficiently solvable in the obvious way. 

The Maximum Weight Induced Matching (MWIM) problem is another example of a problem which can be added to the list of problems above: A set $M$ of edges is an {\em induced matching} in $G$ if the pairwise distance of the edges in $M$ is at least two in $G$. The MWIM problem asks for an induced matching of maximum weight. 
In \cite{BraMos2010}, it is shown that for a hereditary class ${\cal C}$ of graphs, MWIM is solvable in polynomial time if MWIM is solvable in polynomial time on the atoms of ${\cal C}$. This can be applied to (hole,paraglider)-free graphs since for weakly chordal graphs, a polynomial time solution is given in \cite{CamSriTan2003}, and obviously, matched co-bipartite graphs are $3K_2$-free, which means that in such graphs (and in the join of a matched co-bipartite graph and a clique) one has to check only pairs of edges.   

\section{Conclusion}

In this paper we have described the structure of (hole, paraglider)-free atoms (of (hole, diamond)-free atoms, respectively) and some algorithmic consequences. In a forthcoming paper \cite{BerBraGia2011} we will analyze the structure of (hole,diamond)-free graphs and its algorithmic consequences in more detail; in particular, we show that weakly chordal diamond-free atoms are either cliques or chordal bipartite.  

\medskip

There are various other aspects and papers which are related of our work as described subsequently:

\subsection{Related results for subclasses of $P_5$-free graphs}

In \cite{Aleks2004}, Alekseev showed that $P_5$- and paraglider-free atoms are $3K_2$-free which leads to a polynomial time algorithm for the MWIS problem since $3K_2$-free graphs contain at most ${\cal O}(n^4)$ inclusion-maximal independent sets. In \cite{BraLeMah2007}, we improved this result by generalizing the forbidden paraglider subgraph. In \cite{BraHoa2005}, we give a more detailed structural analysis of $P_5$- and paraglider-free atoms. In \cite{BraMos2004}, we describe the structure of prime $P_5$- and co-chair-free graphs and give algorithmic applications. 
The complexity of the MWIS problem for $P_5$-free graphs is an open problem. It is also open for $(P_5,C_5)$-free graphs; such graphs are hole-free. Thus, it is interesting to study subclasses of $P_5$-free graphs (subclasses of $(P_5,C_5)$-free graphs, respectively).  

\subsection{Clique-width} 

In \cite{Brand2004}, we describe the simple structure of ($P_5$,diamond)-free graphs; such graphs can contain $C_5$ and thus, $P_5$- and diamond-free graphs are in general not perfect and incomparable with (hole,diamond)-free graphs. 
($P_5$,diamond)-free graphs have bounded clique-width - see e.g. \cite{CouMakRot2000} for the notion and algorithmic implications of bounded clique-width which has tremendous consequences for efficiently solving hard problems on such graph classes. For the more general class of ($P_5$,gem)-free graphs, the situation is similar: By the Strong Perfect Graph Theorem, (hole,gem)-free graphs are perfect since antiholes with at least seven vertices contain gem. 
The structure of ($P_5$,gem)-free graphs and some algorithmic applications were described in \cite{BodBraKraRaoSpi2005,BraKra2005}. In \cite{BraLeMos2005}, it was shown that ($P_5$,gem)-free graphs have bounded clique-width. 

The clique-width of (hole,diamond)-free graphs, however, is unbounded since e.g. the subclass of chordal bipartite graphs (which are the (hole, triangle)-free graphs), has unbounded clique-width \cite{BraLoz2003}. This illustrates that corresponding subclasses of hole-free graphs are more interesting than those of $P_5$-free graphs.    

\subsection{Open problems}

It would be interesting to describe the structure of (hole,gem)-free graphs. In particular, how can one avoid to use the Strong Perfect Graph Theorem for showing that (hole,gem)-free graphs are perfect?

In \cite{BraGia2011}, we give a polynomial time algorithm for the MWIS problem on hole- and co-chair-free graphs. It would be interesting to obtain better structural results on these graphs. 

\begin{footnotesize}
\renewcommand{\baselinestretch}{0.4}

\end{footnotesize} 

\end{document}